\documentclass[twocolumn]{aastex6}

\usepackage{newtxtext,newtxmath}

\usepackage[T1]{fontenc}
\usepackage{ae,aecompl}

\usepackage{graphicx}	
\usepackage{amsmath}	
\usepackage{amssymb}	
\usepackage{bm}

\newcommand{\cf}{cf.,~}
\newcommand{\ie}{i.e.,~}
\newcommand{\eg}{e.g.,~}

\usepackage{todonotes}
\newcommand{\bhac}{\texttt{BHAC}~}

\shorttitle{Magnetised jets from short GRBs}
\shortauthors{Nathanail et al.}

\begin{document}
\title{On the opening angle of magnetised jets from neutron-star mergers:
  the case of GRB170817A}

\author{Antonios Nathanail\altaffilmark{1}, Ramandeep
  Gill\altaffilmark{2,3}, Oliver Porth\altaffilmark{4}, Christian
  M. Fromm\altaffilmark{1,5}, Luciano Rezzolla\altaffilmark{1,6}}
\altaffiltext{1}{Institut f\"ur Theoretische Physik, Max-von-Laue-Strasse
  1, 60438 Frankfurt, Germany}
\altaffiltext{2}{Department of Physics, The George Washington University, 
  Washington, DC 20052, USA}
\altaffiltext{3}{Department of Natural Sciences, The Open University of 
  Israel, 1 University Road, PO Box 808, Raanana 4353701, Israel}
\altaffiltext{4}{Astronomical Institute Anton Pannekoek, Universeit van Amsterdam, 
Science Park 904, 1098 XH, Amsterdam, The Netherlands}
\altaffiltext{5}{Max-Planck-Institut f\"ur Radioastronomie, Auf dem H\"ugel 69,
D-53121 Bonn, Germany}
\altaffiltext{6}{School of Mathematics, Trinity College, Dublin 2, Ireland}




\label{firstpage}

\begin{abstract}
The observations of GW170817/GRB170817A have confirmed that the
coalescence of a neutron-star binary is the progenitor of a short
gamma-ray burst. In the standard picture of a short gamma-ray burst, a
collimated highly relativistic outflow is launched after merger and it
successfully breaks out from the surrounding ejected matter. Using
initial conditions inspired from numerical-relativity binary neutron-star
merger simulations, we have performed general-relativistic hydrodynamic
(HD) and magnetohydrodynamic (MHD) simulations in which the jet is
launched and propagates self-consistently. The complete set of
simulations suggests that: \textit{(i)} MHD jets have an intrinsic energy
and velocity polar structure with a ``hollow core'' subtending an angle
$\theta_{\rm core}\approx4^{\circ}-5^{\circ}$ and an opening angle of
$\theta_{\rm jet>}\gtrsim10^{\circ}$; \textit{(ii)} MHD jets eject
significant amounts of matter and two orders of magnitude more than HD
jets; \textit{(iii)} the energy stratification in MHD jets naturally
yields the power-law energy scaling
$E(>\Gamma\beta)\propto(\Gamma\beta)^{-4.5}$; \textit{(iv)} MHD jets
provide fits to the afterglow data from GRB170817A that are comparatively
better than those of the HD jets and without free parameters;
\textit{(v)} finally, both of the best-fit HD/MHD models suggest an
observation angle $\theta_{\rm obs} \simeq 21^{\circ}$ for GRB170817A.
\end{abstract}




\section{Introduction}
\label{sec:intro}

The first detection of gravitational waves (GWs) from a binary
neutron-star (BNS) merger, GW170817 \citep{Abbott2017}, was marked by a
coincident detection of a short gamma-ray burst (GRB), GRB170817A
\citep{Savchenko2017,Goldstein2017}. This was followed by observations
across the electromagnetic (EM) spectrum, with the detection of the %
\citep{Abbott2017b} quasi-thermal kilonova emission in UV, optical, and
NIR followed by the delayed detection of the non-thermal afterglow
emission in the X- ($t>8.9\,{\rm\,d}$; \citealt{Troja+17}), optical, and
radio ($t>16.4\,{\rm\,d}$; \citealt{Hallinan2017}) bands.

The continuous brightening of the broadband afterglow flux, with its
peculiar shallow rise ($F_\nu\propto t^{0.8}$) to the peak at $t_{\rm
  pk}\simeq150\,{\rm d}$ post-merger
\citep{Lyman2018,Margutti2018,Mooley2018}, was interpreted using two main
models. The first one considered a ``structured outflow''
\citep[e.g,][]{Gill2018}, namely, a polar-structured jet with a narrow
relativistic core surrounded by low-energy wings
\citep[e.g.,][]{Troja+17,Troja2018,DAvanzo+18,Margutti2018,Lazzati2017c}. 
The second
model considered a ``cocoon'', namely, a wide-angle outflow expanding
quasi-spherically and with radial velocity stratification \citep[e.g.,][]
{Kasliwal2017,Gottlieb2018,Mooley2018}. The subsequent observation of
apparent superluminal motion of the radio flux centroid
\citep{Mooley2018b}, together with the compact size of the radio image
(\ie $\lesssim2\,$mas) \citep{Ghirlanda2019}, strongly favored the
structured jet model as dominating the afterglow emission near and post
$t_{\rm\,pk}$.

\begin{table*}
  \centering
  \begin{tabular}{|l||c|c|c|c|c|c|c|c|c|c|c|c|c|c|}
    \hline
    \hline
    \!\!\!  model&$L$&$t_{\mathrm{inj}}$& $\Gamma_{\mathrm{init}}$&$\theta_{\mathrm{jet}}$&$E_{B_{\phi}}$&
    $ E_{B_{\mathrm{p}}}$& 
    $\frac{E_{B_{\mathrm{p}}}}{E_{B_{\phi}}}$&$\sigma_{\mathrm{max}}$& 
\!\!\!$\beta_{\mathrm{min}}$&$\rho_{\mathrm{max}}$& $a$ & $M_{\mathrm{tot}}$ & $M_{\mathrm{ej}}$ &$\frac{M_{\mathrm{ej}}}{M_{\mathrm{tot}}}$ \\
    \!\!\! &$[\mathrm{erg/s}]$ & $[\mathrm{s}]$ && ${\rm [deg]}$  & $[\mathrm{erg}]$ & $[\mathrm{erg}]$  &
     &&&\!\!\! $[\mathrm{g/cm^3}]$ & & $[M_{\odot}]$& $[M_{\odot}]$& $\%$\\
     &&&&&$10^{49}$&$10^{49}$&&&&$10^{10}$&&&&\\
  \hline
      \! \texttt{HD-tht.6}&$10^{51}$&$0.1$&$10$&
      $6$&$-$&$-$&$-$&$-$&$-$&$1.5 $&
      $0.9375$&$0.108$ &$0.0001$&$0.12$ \\ 
    \! \texttt{HD-tht.3}&$10^{51}$&$0.1$&$10$&
    $3$&$-$&$-$&$-$&$-$&$-$&$1.5 $&
    $0.9375$&$0.108$ &$0.0001$&$0.13$ \\
\! \texttt{MHD-p2t.03} &$-$&$-$&$-$&$-$&$5.0$   & $1.6$  & $0.3$& $0.065$ & $0.13$ & $1.5$ & $0.9375$ & $0.108$& $0.039$& $36.0$ \\
\! \texttt{MHD-p2t.02} &$-$&$-$&$-$&$-$&$10$ & $2.1$ & $0.2$& $0.065$ & $0.13$ & $2.0$ & $0.9375$ & $0.144$& $0.053$&$37.1$\\
\! \texttt{MHD-p2t.12} &$-$&$-$&$-$&$-$&$1.2$ & $1.5$ & $1.2$& $0.036$ & $0.13$ & $1.5$ & $0.9375$ & $0.108$& $0.036$&$34.1$  \\
\hline
\hline
  \end{tabular}  
  \caption{Properties of the various HD and MHD jets considered:
    luminosity of the HD jet ($L$), duration of the HD injection
    ($t_{\mathrm{inj}}$), initial Lorentz factor of the HD jet
    ($\Gamma_{\mathrm{init}}$), initial opening angle of the HD jet
    ($\theta_{\rm jet}$), toroidal and poloidal magnetic energies
    ($E_{B_{\phi}}, E_{B_{\mathrm{p}}}$) and their ratio, maximum
    magnetization in the torus ($\sigma:=B^2/4\pi \rho$), minimum plasma
    parameter in torus ($\beta:=p/p_m$, where $p$ and $p_m$ are the fluid
    and magnetic pressures respectively), maximum density of the torus
    ($\rho_{\mathrm{max}}$) and dimensionless spin parameter of the BH
    ($a:=J/M^2$), initial total mass ($M_{\mathrm{tot}}$), ejected mass
    ($M_{\mathrm{ej}}$) and their ratio.}
  \label{tab:initial}
\end{table*}

Numerical and semi-analytical models of hydrodynamic jets have been
employed to explore the afterglow of GRB170817A and the models that best
fit the afterglow data correspond to structured jets with angular size of
the relativistic core of $\sim 3^{\circ}-5^{\circ}$
\citep{Mooley2018b,Ghirlanda2019,Troja2019}.

Most of the analysis for the outflow of GRB170817A has been done using
semi-analytical models or relativistic hydrodynamic simulations that
launch a jet far from the merger site, with launching radius of
$10^9\,{\rm cm}$. These hydrodynamic studies have been accompanied by
much fewer investigations making use of MHD simulations to study the
properties of such outflows
\citep{Kathirgamaraju2018,Bromberg2018,Geng2019}, and in two cases, the
jets were launched self-consistently via the accretion and rotation of
the black hole \citep{Fernandez2018,Kathirgamaraju2019}. In addition to
such self-consistent evolutions, \citealt{Kathirgamaraju2019} were also
the first to report afterglow lightcurves as derived from the MHD
simulations.

We here report on a series of two-dimensional (2D) general-relativistic
MHD (GRMHD) simulations of jets that are self-consistently launched after
a BNS merger when the merger remnant has collapsed to a black hole
(BH). In addition, we also carry out simulations in general-relativistic
hydrodynamics (HD) -- where the jet is artificially powered via the
injection of energy near the BH -- and use these simulations to compare
and contrast the properties of the MHD and HD jets.

\section{MHD vs HD jets}
\label{sec:comp}

%
\begin{figure*}
  \centering
  \includegraphics[width=0.42\textwidth]{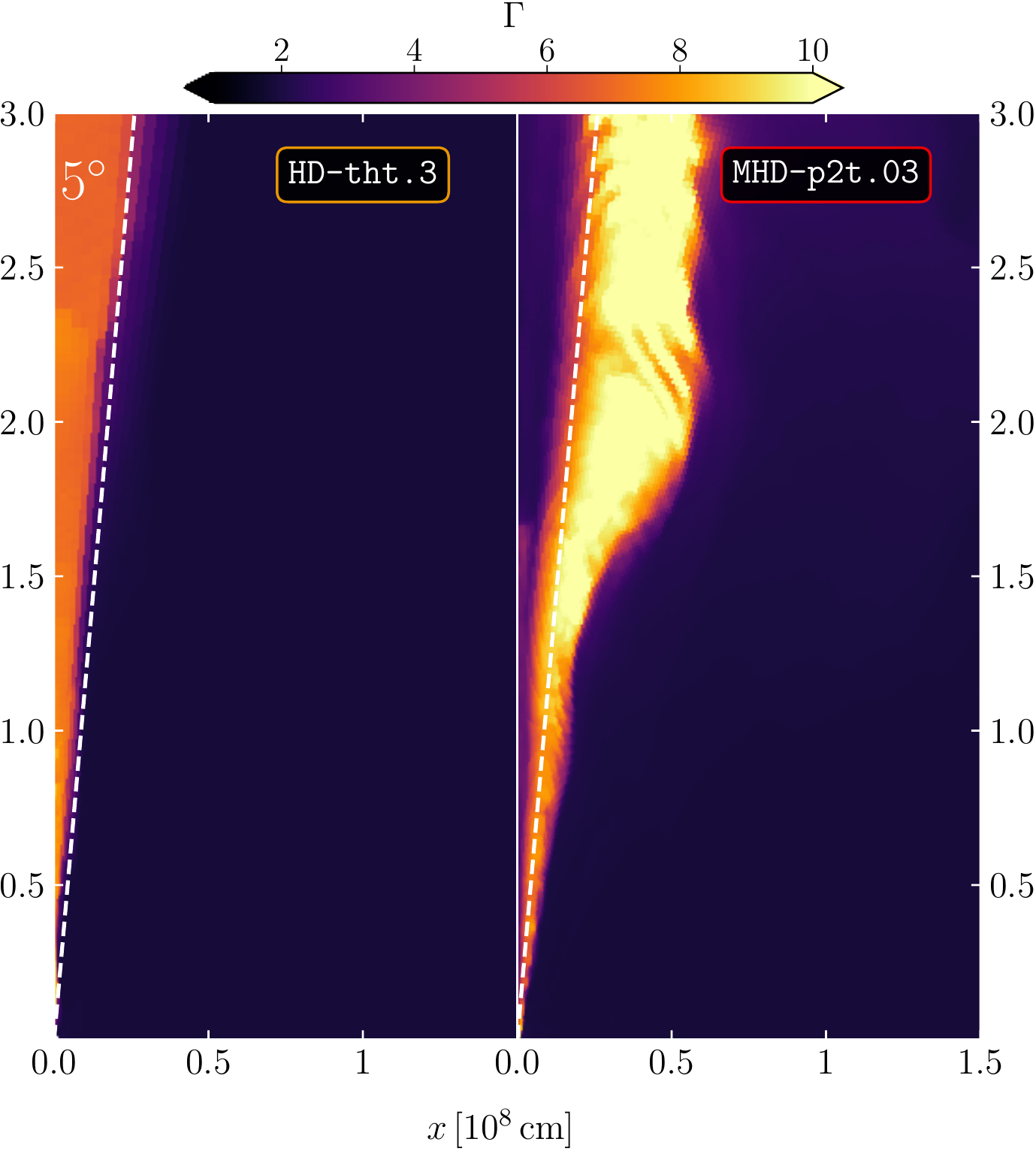}
  \hskip 1.0cm
  \includegraphics[width=0.42\textwidth]{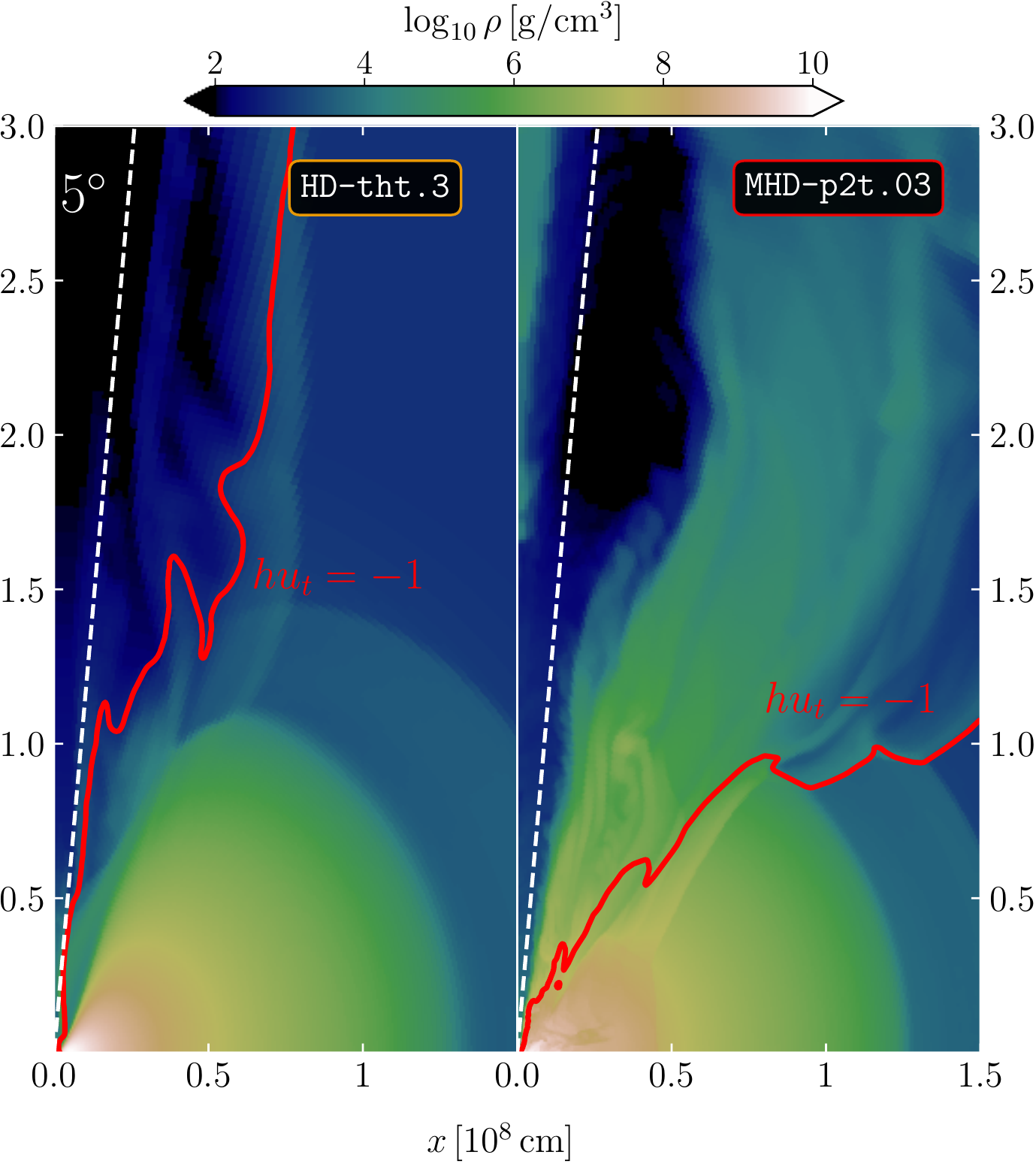}
  \caption{Lorentz factor (left panel) and density (right panel)
    distribution for two representative models: \texttt{MHD-p2t.03} (left
    part of each plot) and \texttt{HD-tht.3} (right part of each plot). The
    dashed white line indicates a cone with opening angle of $5^{\circ}$,
    highlighting the slow core of the MHD jet model. On the right panel
    the red lines denote the contour of $hu_t= -1$, so that matter above
    such line is gravitationally unbound; clearly the amount of ejected
    mass from the MHD jet is significantly larger than in the HD jet
    model.}
  \label{fig:MHDjet}
\end{figure*}
%

We employ \bhac to solve the general-relativistic MHD equations in a Kerr
background spacetime \citep{Porth2017}. In order to describe the ejected
matter and the torus around the compact remnant that was produced after a
BNS merger, we follow the setup introduced in \citet{Nathanail2018c} and
additional information on the numerical setup are reported in the
Appendix. The properties of the models simulated study are listed in
Table~\ref{tab:initial}.

HD jets have been thoroughly studied in the context of short GRBs from
BNS mergers
\citep{Nagakura2014,Murguia-Berthier2014,Murguia-Berthier2016,Duffell2015,Duffell2018}. The
MHD jets in our simulations are launched self-consistently over the
timescale of the simulations, which ranges between $\sim 40\,{\rm ms}$
(for most cases) and $\sim 160\,{\rm ms}$. Overall, the dynamics of the
plasma can be briefly described as follows: starting from a non
self-gravitating torus with initial size $r_{\rm in}=6 M=23.8\,{\rm km}$
and $r_{\rm out} = 14.3 M =56.7\,{\rm km}$ and containing a magnetic
field of various topologies and strengths (\cf Table~\ref{tab:initial}),
the magnetorotational instability (MRI) develops, driving the accretion
of matter and magnetic flux onto the BH. At the same time, the magnetic
pressure in the torus expels the outer layers with an efficiency that
depends strongly on the initial plasma $\beta$ parameter in the torus. As
the MRI saturates and accretion reaches a steady state, the funnel region
above the BH is cleared up and an MHD jet is formed.  This accretion
process can then continue until either the torus is accreted and ejected,
or when the BH has lost much of its reducible energy by spinning down
\citep{Nathanail2016}.

As the MHD jet breaks out from the ejecta that, in our setup, terminate
at a radius of $1,200\,{\rm km}$, it enters in a region of low-density
material where it does not encounter any matter pressure-gradient
that contributed to its collimation. As a result, the jet expands in the
transversal direction while maintaining a high degree of
collimation. More precisely, when the head of the jet reaches $\sim
1,500\,{\rm km}$, the opening angles at a distance of $\sim 500\,{\rm
  km}$ and $\sim 1,500\,{\rm km}$ are $\theta_{\rm jet} \simeq
13^{\circ}$ and $\theta_{\rm jet}\simeq 15^{\circ}$, respectively. By the
time the MHD jet reaches the outer boundary of the computational domain
at $\sim 10,000\,{\rm km}$, the opening angle is still very small and
$\theta_{\rm jet} \simeq 13^{\circ}$. These values depend in detail on
the initial conditions of the jet and on the properties of the ambient
medium \citep{Tchekhovskoy2008}, but do not vary significantly in the
simulations we have considered.

Another robust feature in all our MHD models reported in
Table~\ref{tab:initial}, is an almost hollow core subtending an angle
$\theta_{\rm core}\approx 4^{\circ}-5^{\circ}$, thus much smaller than
the overall opening angle of the MHD jet, $\theta_{\rm jet}\gtrsim
10^{\circ}$; the latter is consistent with numerical-relativity
simulations where the starting point for the launching of such a jet is
reached
\citep{Rezzolla:2011,Kiuchi2014,Dionysopoulou2015,Kawamura2016,Ruiz2016}\footnote{While
  ``hollow core'' is a standard denomination, the core of the jet does
  actually contain matter, but with very small Lorentz factor and
  energy.}. In Fig.~\ref{fig:MHDjet} we show a comparison between an MHD
and a HD jet, where both jets have passed through the torus and the
ejected matter. The Lorentz factor, shown on the left panel, clearly
tends to unity in the inner core of the MHD jet. The appearance of a
hollow cone in MHD jets has been pointed out previously in the literature
\citep{Komissarov2007b,Tchekhovskoy2008,Lyubarsky2009}, but these were
smaller than the one found here in our simulations inspired by BNS merger
scenarios.

\begin{figure*}
  \centering
  \includegraphics[width=0.4\textwidth]{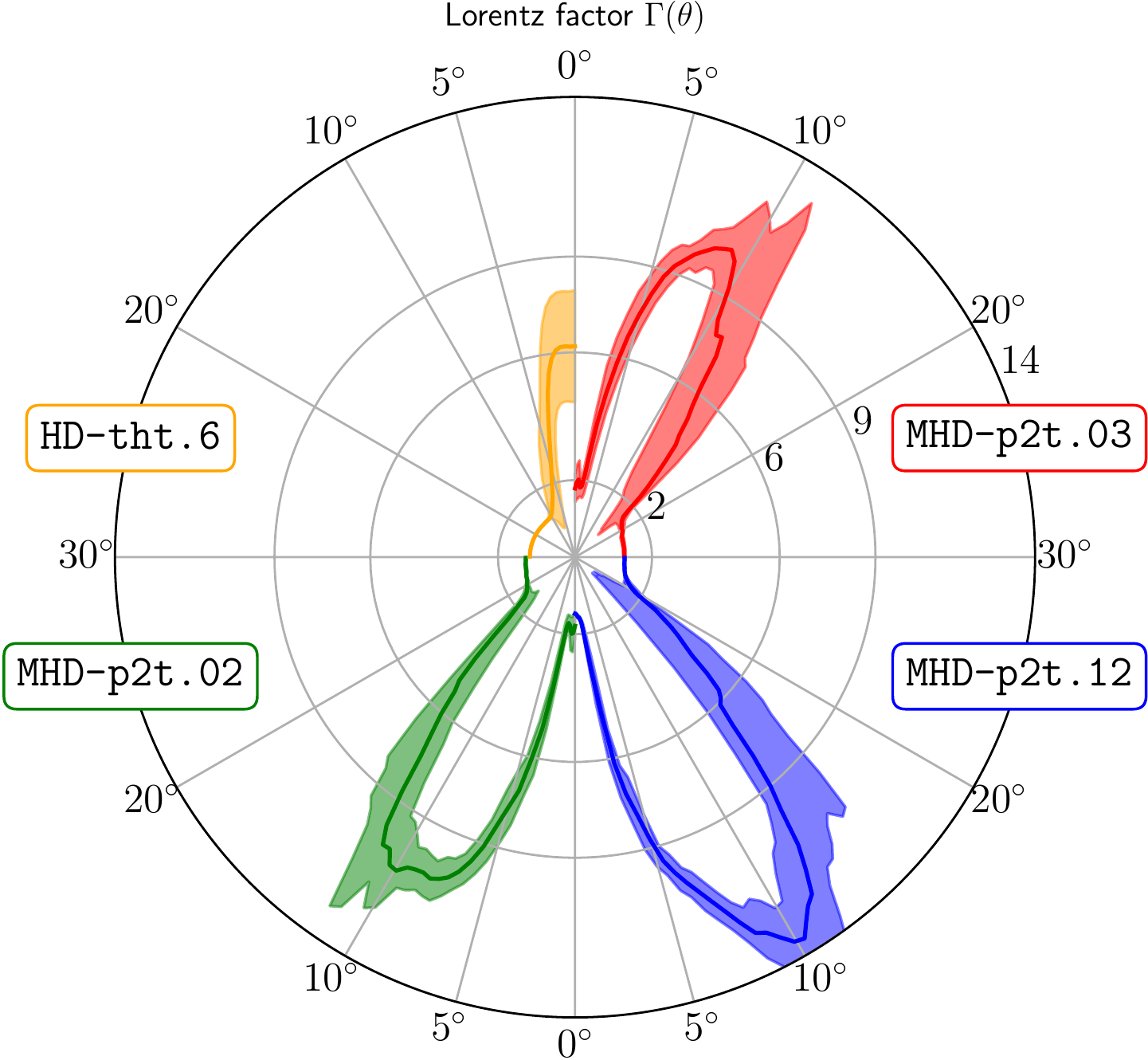}
  \hskip 1.0cm
  \includegraphics[width=0.4\textwidth]{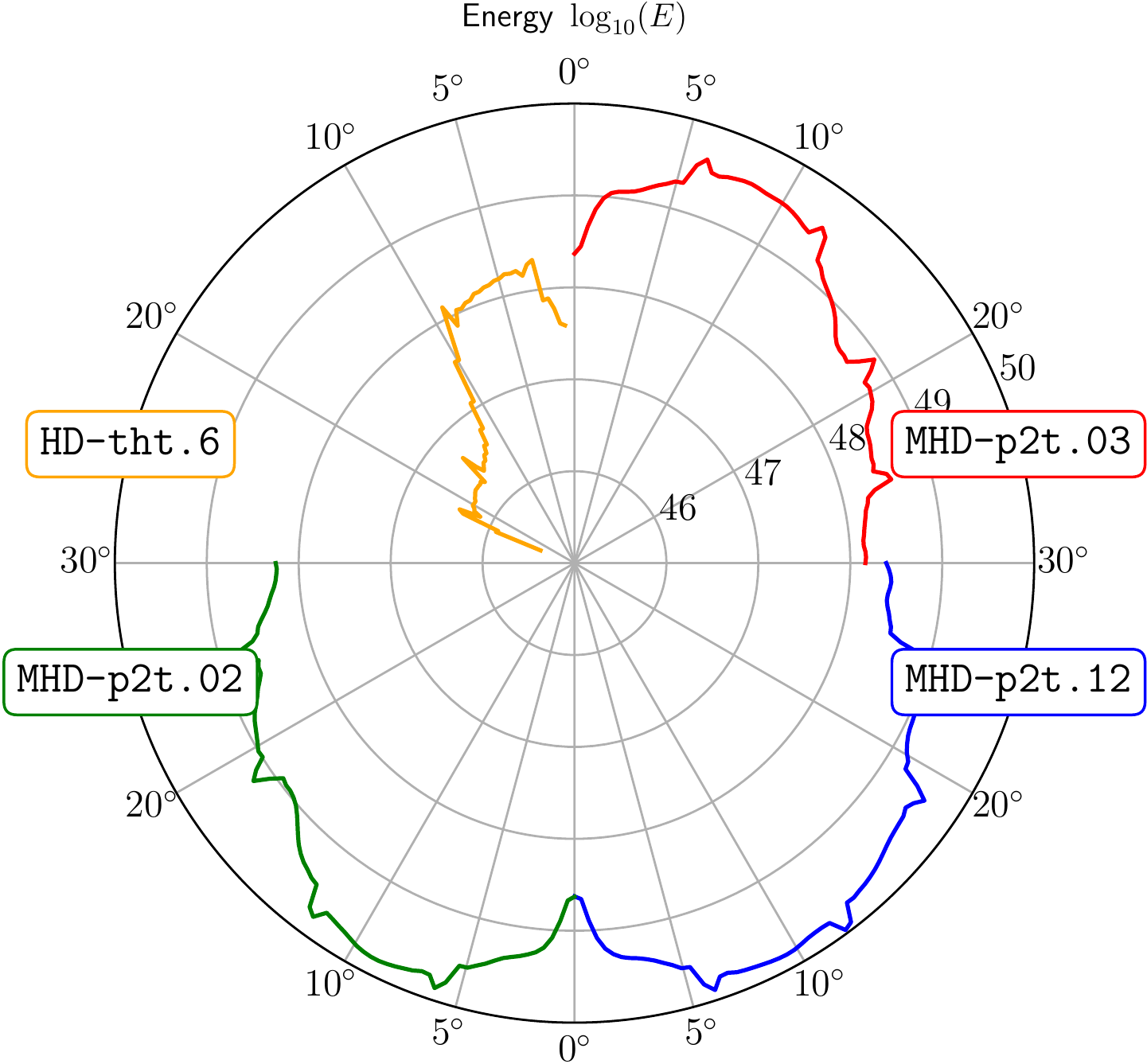}
  \caption{Upper panels: Polar plots of the Lorentz factor for four
    representative outflows over a quadrant (left panel (a)), or within a
    cone of $30^{\circ}$ (right panel (b)); the thick lines show the
    time-averaged values, while the shaded region the $1$-$\sigma$
    variance. Lower panel (c): Polar plot of the energy distribution for
    four representative models within a cone of $30^{\circ}$.}
 \label{fig:ang}
\end{figure*}

The structure and opening angle of the jet models presented in these
studies depend strongly on the collimating agent. In the case of long
GRBs, this agent is represented by the disk wind and the stellar layers
that the jet has to bore. On the other hand, in the case of short GRBs
produced from BNS mergers, once the jet breaks out from the matter
ejected by the merger, it encounters the low-density interstellar medium
(ISM), with number densities $n_{\rm ISM}\sim10^{-3}-10^{-1}~{\rm
  cm}^{-3}$, so that no significant further collimation is expected after
breakout.

\begin{figure*}
  \begin{center}
    \includegraphics[width=0.9\textwidth]{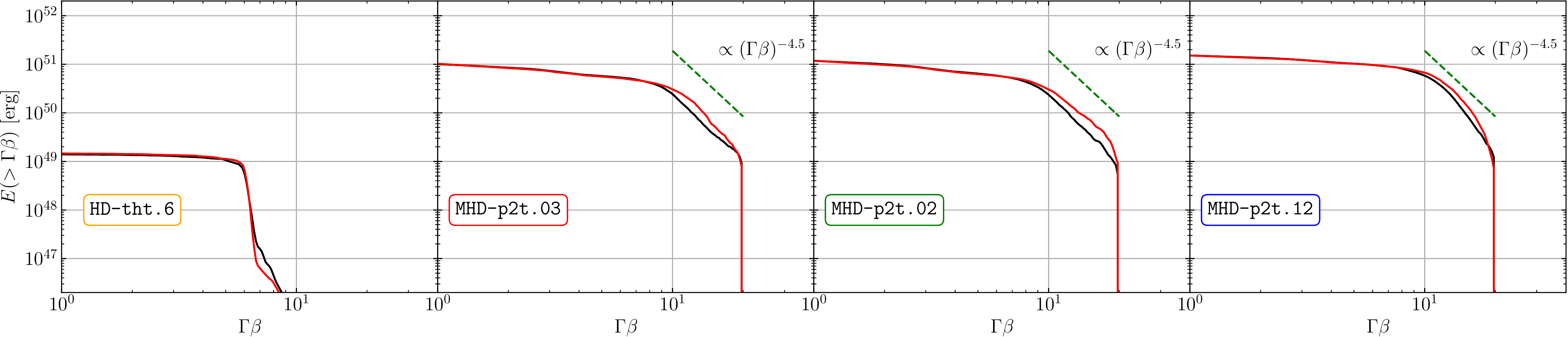}
    \includegraphics[width=0.9\textwidth]{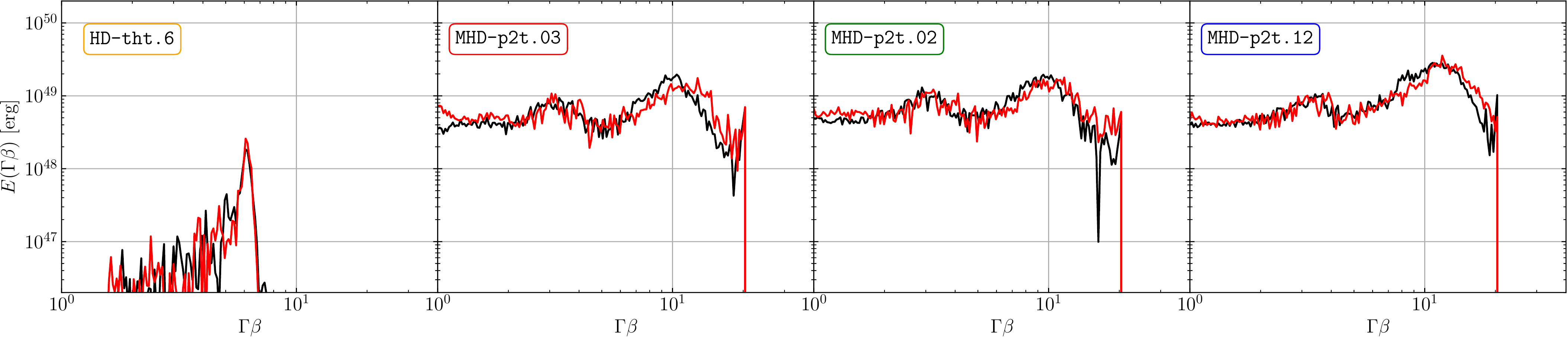}
  \end{center}
  \caption{Energy distributions shown as either as $E=E(>\Gamma\beta)$
    (top panels) or as $E=E(\Gamma\beta)$ (bottom panels) for the four
    representative models. The black and red solid lines represent the
    distribution at different times, $t = 15$ and $20\,{\rm ms}$
    respectively.}
  \label{fig:energy}
\end{figure*}

\cite{Duffell2018} have shown that as a HD jet drills through merger
ejecta, it does not deposit significant energy, and thus has limited
impact on the amount of ejected mass and the appearance of a
kilonova. This is in stark contrast with what happens for MHD jets, the
magnetized torus produces winds, with velocities far below the
relativistic jet but significant enough that a large fraction of the
initial matter distribution becomes unbound. On the right panel of
Fig.~\ref{fig:MHDjet} we show the distribution of the rest-mass density
at time $t\sim26\,{\rm ms}$, after the MHD and the HD jets have broken
out from the merger ejecta.  To quantify how much matter becomes unbound,
we employ the Bernoulli criterion and assume a fluid element to be
unbound if it has a Bernoulli constant $hu_t\leq -1$, where $h$ is the
specific enthalpy of the fluid \citep{Rezzolla_book:2013}. We then apply
this criterion to measure the flux of unbound matter on a 2-sphere of
$4,000\,{\rm km}$ and report in last two columns of
Table~\ref{tab:initial} the amount of ejected mass and the fraction of
the ejected mass with respect to the initial mass of the torus. Note that
in all cases considered the ejected mass is between a few percent of the
initial mass and up to a maximum of $37\%$; furthermore, models with
higher initial $\sigma$, have a larger fraction of unbound matter.

The angular structure of the HD and MHD jets can be better appreciated
through the polar plots in Fig.~\ref{fig:ang}, where we report the
Lorentz factor and the energy, \ie the volume integral up to the outer
boundary of the \emph{total} energy density, relative to the
\textit{unbound} material of three representative MHD jets and of the HD
jet. The Lorentz factor (left panel of Fig.~\ref{fig:ang}) is measured on
a 2-sphere with radius $r\sim2,000\,\mathrm{km}$, and is integrated over
a time interval of $\tau_{\mathrm{avg}}\sim4\,\mathrm{ms}$ to capture
both the variability and the steady features.

Each of the four quadrants refers to one of the models considered, \ie
\texttt{MHD-p2t.03}, \texttt{MHD-p2t.02}, \texttt{MHD-p2t.12}, and
\texttt{HD-tht.6}, with a thick line indicating the time-averaged values
and with the shaded areas showing the $1$-$\sigma$ variance over the time
interval $\tau_{\mathrm{avg}}$, \ie the $68\%$ variation of the Lorentz
factor at each angle. The right panel of Fig.~\ref{fig:ang}, on the other
hand, shows the angular distribution of the energy for the four models,
where the energy is integrated for every angle for
\emph{unbound} matter with $\Gamma >1.2$; such a cut-off is introduced to
avoid the inclusion of comparatively slow material.

In Fig.~\ref{fig:energy} we show instead the energy distribution above a
certain value of $\Gamma\beta$, \ie $E(>\Gamma\beta)$, as a function of
$\Gamma\beta$, both for the HD jet and for the three representative MHD
models. Since the energy $E$ generically grows with $\Gamma\beta$, the
quantity $E(>\Gamma\beta)$ helps capture the nonlinear growth as a
deviation from a constant value and to determine the cut-off at the
highest energies.  The energy is measured after the jet has broken out
from the merger ejecta, \ie $t=10\,{\rm ms}$, and is reported at three
different times with a separation of $5\,\mathrm{ms}$ in time. Note that
the HD jet is less powerful and with an energy that has an almost linear
dependence $\Gamma\beta$ but to $\Gamma\beta \simeq6-7$, when it has a
very sharp fall-of profile at moderate Lorentz factors. Therefore, in a
HD jet a most of the energy is concentrated in the fast-moving material.

On the other hand, all the MHD jets are up to two orders of magnitude
more powerful and have a sub-linear growth of energy with $\Gamma\beta$;
at the same time, the cut-off is less abrupt and preceded by a clear
power-law fall-off at high Lorentz factors, which can be approximated as
$E(>\Gamma\beta)\propto(\Gamma\beta)^{-4.5}$. Hence, in the case of MHD
jets, most of the energy is at $\Gamma\beta \sim 10$, but the energy
distribution in the plasma can reach very large values. Note that a
cut-off of $\Gamma \simeq 20$ is set to avoid to account for portions of
the flow where the accuracy of the numerical solution is reduced because
of the large Lorentz factors reached.

It is worth noting that the bulk of our MHD jets is moving relatively
fast and overall faster than what observed in other simulations
\citep{Gottlieb2018} or analytical modellings
\citep{Mooley2018,Gill2018}, where most of the energy is in slow-moving
material and the power-law behaviour $\Gamma\beta^{-(4-5)}$ is seen
already for $\Gamma\beta\simeq 1$. As a final remark, we note that since
our MHD jets are launched as a result of GRMHD accretion processes, their
energetics cannot be steered from the initial conditions, but is the
self-consistent result of the simulations.

\section{Afterglow emission}
\label{sec:after}

The afterglow emission is expected to be dominated by synchrotron
radiation from electrons at the forward shock propagating into the
low-density ISM and that are accelerated into a power-law energy
distribution of the type $n_e(\Gamma_e)\propto\Gamma_e^{-p}$, where $n_e$
and $\Gamma_e$ are the number density and Lorentz factors of the
electrons, respectively; hereafter, we will assume $p=2.16$, which is
consistent with previous analysis for the afterglow of GRB170817A
\citep{Troja2019,Hajela2019}. Following \cite{Sari1998}, we model the
emission that depends on the microphysical parameters $\epsilon_e$ and
$\epsilon_B$, which describe the fraction of the total internal energy
behind the shock given to electrons and to the magnetic field,
respectively.
The afterglow lightcurves are computed following the angular
distributions of the Lorentz factor and of the energy profile (\cf
Fig.~\ref{fig:ang}), together with the energy distribution in
$\Gamma\beta$ (\cf~\ref{fig:energy}).  The angular structure is binned
uniformly in $200$ angles along the $\theta$ direction, which yields the
initial $\Gamma_0(\theta)$ and isotropic-equivalent energy $E_{\rm
  iso}(\theta)$ of the flow \citep[see][for
  details]{Granot1999,Gill2018}.

\begin{figure}
  \includegraphics[width=0.45\textwidth]{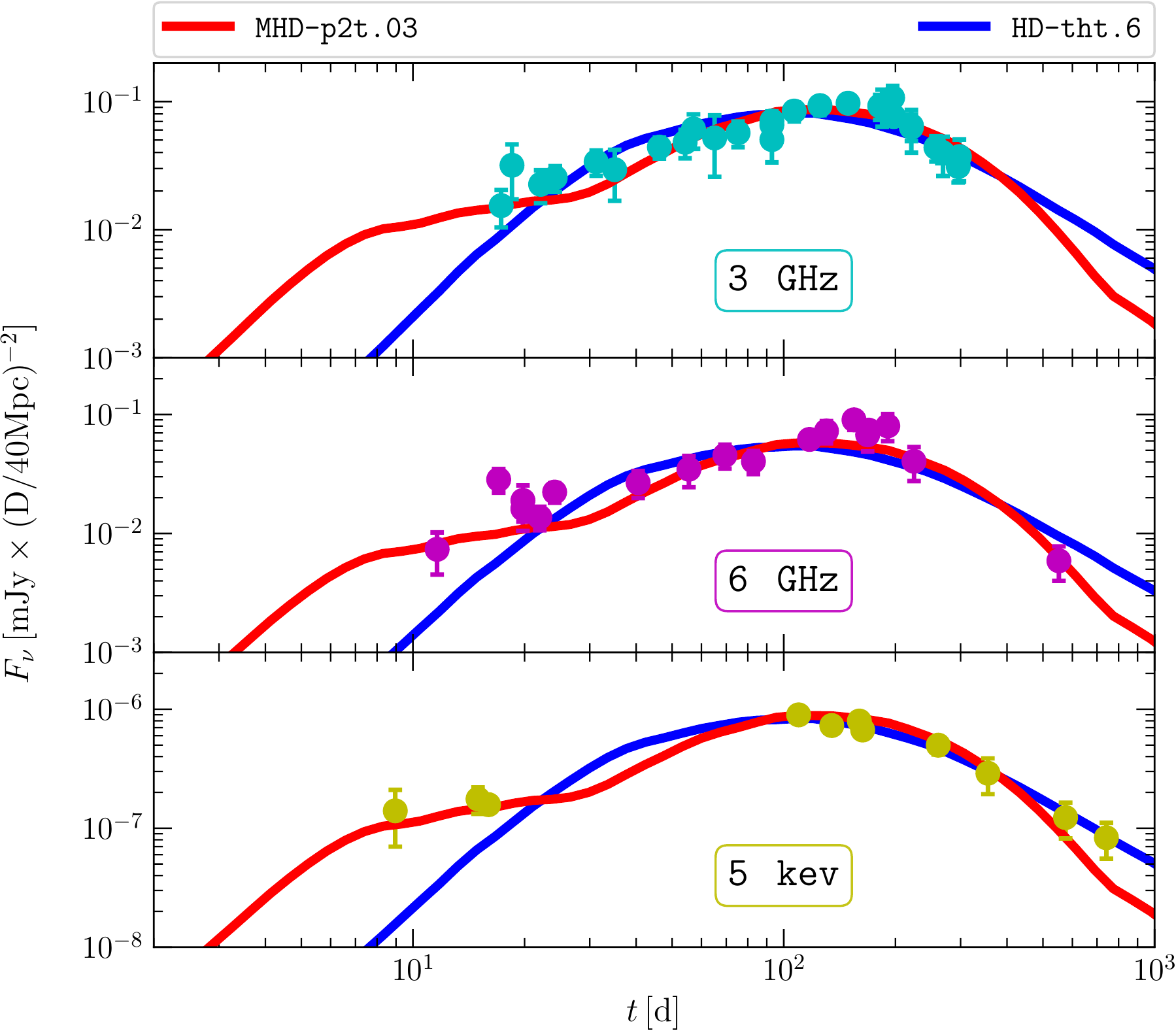}
  \caption{Broad band observations of GRB170817A with the best-fit lightcurves
    of models \texttt{MHD-p2t.03} (red line; see main text for the
    fitting parameters) and \texttt{HD-tht.6} (dashed blue line; see main
    text for the fitting parameters).}
	\label{fig:aft}
\end{figure}

As representative examples of our fits, we make use of model
\texttt{HD-tht.6} and model \texttt{MHD-p2t.03}. For the data, on the
other hand, we employ the most recent afterglow data, \ie
$t\lesssim743\,{\rm d}$ after merger \citep[see, \eg][for the latest
  observations in X-rays]{Hajela2019} consisting of X-ray emission at
$5\,$keV and VLA radio observations at $3$ and $6\,$GHz
\citep{Margutti2017,Margutti2018,Alexander2017,Alexander2018,Hallinan2017,Mooley2018,Mooley2018c,Dobie2018,Troja2018,Troja2019,Hajela2019}. The
fit is performed with five free parameters, namely: the observer angle
$\theta_{\rm obs}$, the energy of the burst $E$, the fraction of the
total energy in the electrons $\epsilon_e$, the fraction of the total
energy in the magnetic field $\epsilon_B$, and the circum-merger density,
$n_{\rm ISM}$. Note that the parameter space is degenerate since the
model parameters outnumber the available constraints from the data
\citep[see, \eg][]{Gill+19}. The best-fit parameters are then found using
a genetic algorithm to optimize the parameter selection and minimize the
reduced $\chi^2_{\nu}$ \citep{Fromm2019}, while the fitting procedure is
applied simultaneously to the three different bands.

The afterglow lightcurves relative to the set of parameters providing the
best fits for the two models \texttt{MHD-p2t.03} and \texttt{HD-tht.6},
along with the observational data, are shown in Fig.~\ref{fig:aft} for a
source at $40\,{\rm Mpc}$, where the upper and middle panels correspond
to radio observations at $3\,$ and $6\,$GHz, while the lower panel to
X-ray observations at $5\,$keV.

Overall, the MHD jet model \texttt{MHD-p2t.03} yields a better fit to the
data, with a reduced $\chi^2_{\nu} = 2.5$ and parameters $\theta_{\rm
  obs}=21.5^{\circ},\,E= 10^{50.85}\,\mathrm{erg},\,
\log_{10}(\epsilon_e)={-0.99},\, \log_{10}(\epsilon_B)={-4.4},\,$ and
$n_{\rm ISM}=10^{-2.04}\,\mathrm{cm^3}$ (red line). It captures well the
first data points in the afterglow, together with the peak and the
fall-off.  The HD jet model \texttt{HD-tht.6}, on the other hand,
provides a less-good fit with reduced chi-squared are $\chi^2_{\nu} =
4.04$ and parameters $\theta_{\rm obs}=21.4^{\circ},\,E=
10^{51.01}\,\mathrm{erg},\, \log_{10}(\epsilon_e)={-0.27},\,
\log_{10}(\epsilon_B)={-2.8},\,$ and $n_{\rm
  ISM}=10^{-4.14}\,\mathrm{cm^3}$ (dashed blue line); however, it also
yields a better match to the very late decay in the X-ray emission
till $743$ days after the merger (model \texttt{HD-tht.3} has
$\chi^2_{\nu} = 5.06$ and an HD jet with $\theta_{\rm jet}=16^{\circ}$
has even larger reduced chi-squared). 
Interestingly, both of the best-fit models
suggest an observation angle $\theta_{\rm obs} \simeq 21^{\circ}$, which
can then be taken as a robust feature of the emission of GRB170817A. Our
estimates are thus consistent with those of \cite{Mooley2018b,Troja2019}, 
and smaller than those coming from the
semi-analytical and analytical modelings, which suggest instead
$\theta_{\rm obs} \simeq 30^{\circ}$ \citep{Hajela2019}.

It is worth noting that when all the physical parameters -- \ie $E$,
$\epsilon_e$, $\epsilon_B$, and $n_{\rm ISM}$ -- are kept the same, the
HD/MHD light curves show a marked difference. Indeed, while both
lightcurves have similar power-law rise and fall-offs, the evolution of
peak-times are considerably different, with the HD having a monotonic
dependence of the peak-times with the viewing angle, with peak-times
increasing as viewing angles become larger. The MHD
lightcurves, instead, do not have a minimum peak-time at the smallest
viewing angle, but for $\theta_{\rm obs} \gtrsim \theta_{\rm core}$; the
peak-time then increases steeply as the viewing angle grows. This
considerable difference between the two afterglow lightcurves disappears
for larger angles, that is, when the jets are observed off-axis.

\section{Conclusions}
\label{sec:con}

We have performed a number of general-relativistic HD and MHD simulations
to model the launching of a jet after a BNS merger and contrast the
dynamics and appearance of HD and MHD jets. Overall, we find that:\\
\textit{(i)} MHD jets have an intrinsic energy and velocity structure in
the polar direction characterised by a ``hollow core'' subtending an
angle $\theta_{\rm core}\approx4^{\circ}-5^{\circ}$ and an opening angle
of $\theta_{\rm jet>}\gtrsim10^{\circ}$. HD jets, on the other hand, have
a uniform energy and polar structure and much smaller opening angles of
$\theta_{\rm jet}\sim3^{\circ}$.
\textit{(ii)} MHD jets eject significant amounts of matter, amounting to
$\lesssim30\%$ of the total mass of the system and about two orders of
magnitude more than HD jets.
\textit{(iii)} The energy stratification in MHD jets naturally yields the
power-law energy scaling $E(>\Gamma\beta)\propto(\Gamma\beta)^{-4.5}$
often introduced in analytical modelling. This feature is robust and does
not require special tuning as is the case instead for HD jets.
\textit{(iv)} MHD jets provide fits to the afterglow data from GRB170817A
in three different bands ($3\,$GHz, $6\,$GHz and $5\,$keV) that are not
only very good but also comparatively better than those of the HD
jets. While even better fits can be constructed with suitably constructed
HD jets, the fit obtained with MHD jets is robust and without free
parameters.
\textit{(v)} Both of the best-fit HD/MHD models suggest an observation
angle $\theta_{\rm obs} \simeq 21^{\circ}$ for GRB170817A.

While this is arguably the most comprehensive exploration of jet
launching from BNS mergers, explore and contrasting for the first time HD
and MHD jets, future work will have to include additional jet models, a
closer comparison with other models proposed in the literature, and a
step towards imaging in the radio band. \\

\smallskip\noindent\textit{Acknowledgements.~}Support comes in part also
from ``PHAROS'', COST Action CA16214 and the LOEWE-Program in HIC for
FAIR. The simulations were performed on the SuperMUC cluster at the LRZ
in Garching, on the LOEWE cluster at the CSC in Frankfurt, and on the
HazelHen cluster at the HLRS in Stuttgart.




\bibliographystyle{mnras}
\bibliography{aeireferences,refs}

\begin{thebibliography}{}
\makeatletter
\relax
\def\mn@urlcharsother{\let\do\@makeother \do\$\do\&\do\#\do\^\do\_\do\%\do\~}
\def\mn@doi{\begingroup\mn@urlcharsother \@ifnextchar [ {\mn@doi@}
  {\mn@doi@[]}}
\def\mn@doi@[#1]#2{\def\@tempa{#1}\ifx\@tempa\@empty \href
  {http://dx.doi.org/#2} {doi:#2}\else \href {http://dx.doi.org/#2} {#1}\fi
  \endgroup}
\def\mn@eprint#1#2{\mn@eprint@#1:#2::\@nil}
\def\mn@eprint@arXiv#1{\href {http://arxiv.org/abs/#1} {{\tt arXiv:#1}}}
\def\mn@eprint@dblp#1{\href {http://dblp.uni-trier.de/rec/bibtex/#1.xml}
  {dblp:#1}}
\def\mn@eprint@#1:#2:#3:#4\@nil{\def\@tempa {#1}\def\@tempb {#2}\def\@tempc
  {#3}\ifx \@tempc \@empty \let \@tempc \@tempb \let \@tempb \@tempa \fi \ifx
  \@tempb \@empty \def\@tempb {arXiv}\fi \@ifundefined
  {mn@eprint@\@tempb}{\@tempb:\@tempc}{\expandafter \expandafter \csname
  mn@eprint@\@tempb\endcsname \expandafter{\@tempc}}}

\bibitem[\protect\citeauthoryear{{Abramowicz}, {Jaroszynski}  \&
  {Sikora}}{{Abramowicz} et~al.}{1978}]{Abramowicz78}
{Abramowicz} M.,  {Jaroszynski} M.,   {Sikora} M.,  1978, Astron. Astrophys.,
  \href {http://adsabs.harvard.edu/abs/1978A%26A....63..221A} {63, 221}

\bibitem[\protect\citeauthoryear{{Alexander} et~al.,}{{Alexander}
  et~al.}{2017}]{Alexander2017}
{Alexander} K.~D.,  et~al., 2017, \mn@doi [Astrophys. J. Letters]
  {10.3847/2041-8213/aa905d}, \href
  {http://adsabs.harvard.edu/abs/2017ApJ...848L..21A} {848, L21}

\bibitem[\protect\citeauthoryear{{Alexander} et~al.,}{{Alexander}
  et~al.}{2018}]{Alexander2018}
{Alexander} K.~D.,  et~al., 2018, \mn@doi [Astrophys. J. Letters]
  {10.3847/2041-8213/aad637}, \href
  {http://adsabs.harvard.edu/abs/2018ApJ...863L..18A} {863, L18}

\bibitem[\protect\citeauthoryear{{Bovard}, {Martin}, {Guercilena}, {Arcones},
  {Rezzolla}  \& {Korobkin}}{{Bovard} et~al.}{2017}]{Bovard2017}
{Bovard} L.,  {Martin} D.,  {Guercilena} F.,  {Arcones} A.,  {Rezzolla} L.,
  {Korobkin} O.,  2017, Phys. Rev. D, \href
  {http://adsabs.harvard.edu/abs/2017arXiv170909630B} {96, 124005}

\bibitem[\protect\citeauthoryear{{Bromberg}, {Tchekhovskoy}, {Gottlieb},
  {Nakar}  \& {Piran}}{{Bromberg} et~al.}{2018}]{Bromberg2018}
{Bromberg} O.,  {Tchekhovskoy} A.,  {Gottlieb} O.,  {Nakar} E.,   {Piran} T.,
  2018, \mn@doi [Mon. Not. R. Astron. Soc.] {10.1093/mnras/stx3316}, \href
  {http://adsabs.harvard.edu/abs/2018MNRAS.475.2971B} {475, 2971}

\bibitem[\protect\citeauthoryear{{D'Avanzo} et~al.,}{{D'Avanzo}
  et~al.}{2018}]{DAvanzo+18}
{D'Avanzo} P.,  et~al., 2018, \mn@doi [\aap] {10.1051/0004-6361/201832664},
  \href {http://adsabs.harvard.edu/abs/2018A%26A...613L...1D} {613, L1}

\bibitem[\protect\citeauthoryear{{Dietrich}, {Ujevic}, {Tichy}, {Bernuzzi}  \&
  {Br{\"u}gmann}}{{Dietrich} et~al.}{2017}]{Dietrich2017}
{Dietrich} T.,  {Ujevic} M.,  {Tichy} W.,  {Bernuzzi} S.,   {Br{\"u}gmann} B.,
  2017, \mn@doi [Phys. Rev. D] {10.1103/PhysRevD.95.024029}, \href
  {http://adsabs.harvard.edu/abs/2017PhRvD..95b4029D} {95, 024029}

\bibitem[\protect\citeauthoryear{{Dionysopoulou}, {Alic}  \&
  {Rezzolla}}{{Dionysopoulou} et~al.}{2015}]{Dionysopoulou2015}
{Dionysopoulou} K.,  {Alic} D.,   {Rezzolla} L.,  2015, \mn@doi [Phys. Rev. D]
  {10.1103/PhysRevD.92.084064}, \href
  {http://adsabs.harvard.edu/abs/2015PhRvD..92h4064D} {92, 084064}

\bibitem[\protect\citeauthoryear{{Dobie} et~al.,}{{Dobie}
  et~al.}{2018}]{Dobie2018}
{Dobie} D.,  et~al., 2018, \mn@doi [Astrophys. J. Letters]
  {10.3847/2041-8213/aac105}, \href
  {http://adsabs.harvard.edu/abs/2018ApJ...858L..15D} {858, L15}

\bibitem[\protect\citeauthoryear{{Duffell}, {Quataert}  \&
  {MacFadyen}}{{Duffell} et~al.}{2015}]{Duffell2015}
{Duffell} P.~C.,  {Quataert} E.,   {MacFadyen} A.~I.,  2015, \mn@doi
  [Astrophys. J.] {10.1088/0004-637X/813/1/64}, \href
  {http://adsabs.harvard.edu/abs/2015ApJ...813...64D} {813, 64}

\bibitem[\protect\citeauthoryear{{Duffell}, {Quataert}, {Kasen}  \&
  {Klion}}{{Duffell} et~al.}{2018}]{Duffell2018}
{Duffell} P.~C.,  {Quataert} E.,  {Kasen} D.,   {Klion} H.,  2018, preprint,
  \href {http://adsabs.harvard.edu/abs/2018arXiv180610616D} {} (\mn@eprint
  {arXiv} {1806.10616})

\bibitem[\protect\citeauthoryear{{Fern{\'a}ndez}, {Tchekhovskoy}, {Quataert},
  {Foucart}  \& {Kasen}}{{Fern{\'a}ndez} et~al.}{2018}]{Fernandez2018}
{Fern{\'a}ndez} R.,  {Tchekhovskoy} A.,  {Quataert} E.,  {Foucart} F.,
  {Kasen} D.,  2018, \mn@doi [Mon. Not. R. Astron. Soc.]
  {10.1093/mnras/sty2932}, \href
  {http://adsabs.harvard.edu/abs/2018MNRAS.tmp.2798F} {}

\bibitem[\protect\citeauthoryear{{Fishbone} \& {Moncrief}}{{Fishbone} \&
  {Moncrief}}{1976}]{Fishbone76}
{Fishbone} L.~G.,  {Moncrief} V.,  1976, Astrophys. J., \href
  {http://adsabs.harvard.edu/abs/1976ApJ...207..962F} {207, 962}

\bibitem[\protect\citeauthoryear{{Foucart}, {O'Connor}, {Roberts}, {Kidder},
  {Pfeiffer}  \& {Scheel}}{{Foucart} et~al.}{2016}]{Foucart2016a}
{Foucart} F.,  {O'Connor} E.,  {Roberts} L.,  {Kidder} L.~E.,  {Pfeiffer}
  H.~P.,   {Scheel} M.~A.,  2016, \mn@doi [Phys. Rev. D]
  {10.1103/PhysRevD.94.123016}, \href
  {http://adsabs.harvard.edu/abs/2016PhRvD..94l3016F} {94, 123016}

\bibitem[\protect\citeauthoryear{{Fromm} et~al.,}{{Fromm}
  et~al.}{2019}]{Fromm2019}
{Fromm} C.~M.,  et~al., 2019, \mn@doi [Astron. Astrophys.]
  {10.1051/0004-6361/201834724}, \href
  {https://ui.adsabs.harvard.edu/abs/2019A&A...629A...4F} {629, A4}

\bibitem[\protect\citeauthoryear{{Fujibayashi}, {Kiuchi}, {Nishimura},
  {Sekiguchi}  \& {Shibata}}{{Fujibayashi} et~al.}{2018}]{Fujibayashi2017b}
{Fujibayashi} S.,  {Kiuchi} K.,  {Nishimura} N.,  {Sekiguchi} Y.,   {Shibata}
  M.,  2018, \mn@doi [Astrophys. J.] {10.3847/1538-4357/aabafd}, \href
  {http://adsabs.harvard.edu/abs/2018ApJ...860...64F} {860, 64}

\bibitem[\protect\citeauthoryear{{Geng}, {Zhang}, {K{\"o}lligan}, {Kuiper}  \&
  {Huang}}{{Geng} et~al.}{2019}]{Geng2019}
{Geng} J.-J.,  {Zhang} B.,  {K{\"o}lligan} A.,  {Kuiper} R.,   {Huang} Y.-F.,
  2019, \mn@doi [Astrophys. J. Letters] {10.3847/2041-8213/ab224b}, \href
  {https://ui.adsabs.harvard.edu/abs/2019ApJ...877L..40G} {877, L40}

\bibitem[\protect\citeauthoryear{{Ghirlanda} et~al.,}{{Ghirlanda}
  et~al.}{2019}]{Ghirlanda2019}
{Ghirlanda} G.,  et~al., 2019, \mn@doi [Science] {10.1126/science.aau8815},
  \href {https://ui.adsabs.harvard.edu/abs/2019Sci...363..968G} {363, 968}

\bibitem[\protect\citeauthoryear{{Gill} \& {Granot}}{{Gill} \&
  {Granot}}{2018}]{Gill2018}
{Gill} R.,  {Granot} J.,  2018, \mn@doi [Mon. Not. R. Astron. Soc.]
  {10.1093/mnras/sty1214}, \href
  {http://adsabs.harvard.edu/abs/2018MNRAS.tmp.1159G} {}

\bibitem[\protect\citeauthoryear{{Gill}, {Granot}, {De Colle}  \&
  {Urrutia}}{{Gill} et~al.}{2019}]{Gill+19}
{Gill} R.,  {Granot} J.,  {De Colle} F.,   {Urrutia} G.,  2019,
  arXiv:1902.10303, e-prints, \href
  {https://ui.adsabs.harvard.edu/abs/2019arXiv190210303G} {}

\bibitem[\protect\citeauthoryear{{Goldstein} et~al.,}{{Goldstein}
  et~al.}{2017}]{Goldstein2017}
{Goldstein} A.,  et~al., 2017, \mn@doi [Astrophys. J. Letters]
  {10.3847/2041-8213/aa8f41}, \href
  {http://adsabs.harvard.edu/abs/2017ApJ...848L..14G} {848, L14}

\bibitem[\protect\citeauthoryear{{Gottlieb}, {Nakar}  \& {Piran}}{{Gottlieb}
  et~al.}{2018}]{Gottlieb2018}
{Gottlieb} O.,  {Nakar} E.,   {Piran} T.,  2018, \mn@doi [Mon. Not. R. Astron.
  Soc.] {10.1093/mnras/stx2357}, \href
  {http://adsabs.harvard.edu/abs/2018MNRAS.473..576G} {473, 576}

\bibitem[\protect\citeauthoryear{{Granot}, {Piran}  \& {Sari}}{{Granot}
  et~al.}{1999}]{Granot1999}
{Granot} J.,  {Piran} T.,   {Sari} R.,  1999, \mn@doi [Astrophys. J.]
  {10.1086/306884}, \href
  {https://ui.adsabs.harvard.edu/abs/1999ApJ...513..679G} {513, 679}

\bibitem[\protect\citeauthoryear{{Hajela} et~al.,}{{Hajela}
  et~al.}{2019}]{Hajela2019}
{Hajela} A.,  et~al., 2019, \mn@doi [Astrophys. J. Lett.]
  {10.3847/2041-8213/ab5226}, \href
  {https://ui.adsabs.harvard.edu/abs/2019ApJ...886L..17H} {886, L17}

\bibitem[\protect\citeauthoryear{{Hallinan} et~al.,}{{Hallinan}
  et~al.}{2017}]{Hallinan2017}
{Hallinan} G.,  et~al., 2017, \mn@doi [Science] {10.1126/science.aap9855},
  \href {http://adsabs.harvard.edu/abs/2017Sci...358.1579H} {358, 1579}

\bibitem[\protect\citeauthoryear{{Kasliwal} et~al.,}{{Kasliwal}
  et~al.}{2017}]{Kasliwal2017}
{Kasliwal} M.~M.,  et~al., 2017, \mn@doi [Science] {10.1126/science.aap9455},
  \href {http://adsabs.harvard.edu/abs/2017Sci...358.1559K} {358, 1559}

\bibitem[\protect\citeauthoryear{{Kathirgamaraju}, {Barniol Duran}  \&
  {Giannios}}{{Kathirgamaraju} et~al.}{2018}]{Kathirgamaraju2018}
{Kathirgamaraju} A.,  {Barniol Duran} R.,   {Giannios} D.,  2018, \mn@doi [Mon.
  Not. R. Astron. Soc.] {10.1093/mnrasl/slx175}, \href
  {http://adsabs.harvard.edu/abs/2018MNRAS.473L.121K} {473, L121}

\bibitem[\protect\citeauthoryear{{Kathirgamaraju}, {Tchekhovskoy}, {Giannios}
  \& {Barniol Duran}}{{Kathirgamaraju} et~al.}{2019}]{Kathirgamaraju2019}
{Kathirgamaraju} A.,  {Tchekhovskoy} A.,  {Giannios} D.,   {Barniol Duran} R.,
  2019, \mn@doi [Mon. Not. R. Astron. Soc.] {10.1093/mnrasl/slz012}, \href
  {https://ui.adsabs.harvard.edu/abs/2019MNRAS.484L..98K} {484, L98}

\bibitem[\protect\citeauthoryear{{Kawamura}, {Giacomazzo}, {Kastaun}, {Ciolfi},
  {Endrizzi}, {Baiotti}  \& {Perna}}{{Kawamura} et~al.}{2016}]{Kawamura2016}
{Kawamura} T.,  {Giacomazzo} B.,  {Kastaun} W.,  {Ciolfi} R.,  {Endrizzi} A.,
  {Baiotti} L.,   {Perna} R.,  2016, \mn@doi [Phys. Rev. D]
  {10.1103/PhysRevD.94.064012}, \href
  {http://adsabs.harvard.edu/abs/2016PhRvD..94f4012K} {94, 064012}

\bibitem[\protect\citeauthoryear{{Kiuchi}, {Kyutoku}, {Sekiguchi}, {Shibata}
  \& {Wada}}{{Kiuchi} et~al.}{2014}]{Kiuchi2014}
{Kiuchi} K.,  {Kyutoku} K.,  {Sekiguchi} Y.,  {Shibata} M.,   {Wada} T.,  2014,
  \mn@doi [Phys. Rev. D] {10.1103/PhysRevD.90.041502}, \href
  {http://adsabs.harvard.edu/abs/2014PhRvD..90d1502K} {90, 041502}

\bibitem[\protect\citeauthoryear{{Kiuchi}, {Kyutoku}, {Sekiguchi}  \&
  {Shibata}}{{Kiuchi} et~al.}{2018}]{Kiuchi2017}
{Kiuchi} K.,  {Kyutoku} K.,  {Sekiguchi} Y.,   {Shibata} M.,  2018, \mn@doi
  [Phys. Rev. D] {10.1103/PhysRevD.97.124039}, \href
  {http://adsabs.harvard.edu/abs/2018PhRvD..97l4039K} {97, 124039}

\bibitem[\protect\citeauthoryear{{Komissarov}, {Barkov}, {Vlahakis}  \&
  {K{\"o}nigl}}{{Komissarov} et~al.}{2007}]{Komissarov2007b}
{Komissarov} S.~S.,  {Barkov} M.~V.,  {Vlahakis} N.,   {K{\"o}nigl} A.,  2007,
  \mn@doi [Mon. Not. R. Astron. Soc.] {10.1111/j.1365-2966.2007.12050.x}, \href
  {http://adsabs.harvard.edu/abs/2007MNRAS.380...51K} {380, 51}

\bibitem[\protect\citeauthoryear{{Lazzati}, {Perna}, {Morsony}, {Lopez-Camara},
  {Cantiello}, {Ciolfi}, {Giacomazzo}  \& {Workman}}{{Lazzati}
  et~al.}{2018}]{Lazzati2017c}
{Lazzati} D.,  {Perna} R.,  {Morsony} B.~J.,  {Lopez-Camara} D.,  {Cantiello}
  M.,  {Ciolfi} R.,  {Giacomazzo} B.,   {Workman} J.~C.,  2018, \mn@doi [Phys.
  Rev. Lett.] {10.1103/PhysRevLett.120.241103}, \href
  {https://ui.adsabs.harvard.edu/abs/2018PhRvL.120x1103L} {120, 241103}

\bibitem[\protect\citeauthoryear{{Lyman} et~al.,}{{Lyman}
  et~al.}{2018}]{Lyman2018}
{Lyman} J.~D.,  et~al., 2018, \mn@doi [Nature Astronomy]
  {10.1038/s41550-018-0511-3}, \href
  {http://adsabs.harvard.edu/abs/2018NatAs.tmp...88L} {}

\bibitem[\protect\citeauthoryear{{Lyubarsky}}{{Lyubarsky}}{2009}]{Lyubarsky2009}
{Lyubarsky} Y.,  2009, \mn@doi [Astrophys. J.] {10.1088/0004-637X/698/2/1570},
  \href {https://ui.adsabs.harvard.edu/abs/2009ApJ...698.1570L} {698, 1570}

\bibitem[\protect\citeauthoryear{{Margutti} et~al.,}{{Margutti}
  et~al.}{2017}]{Margutti2017}
{Margutti} R.,  et~al., 2017, \mn@doi [Astrophys. J. Letters]
  {10.3847/2041-8213/aa9057}, \href
  {http://adsabs.harvard.edu/abs/2017ApJ...848L..20M} {848, L20}

\bibitem[\protect\citeauthoryear{{Margutti} et~al.,}{{Margutti}
  et~al.}{2018}]{Margutti2018}
{Margutti} R.,  et~al., 2018, \mn@doi [Astrophys. J. Letters]
  {10.3847/2041-8213/aab2ad}, \href
  {http://adsabs.harvard.edu/abs/2018ApJ...856L..18M} {856, L18}

\bibitem[\protect\citeauthoryear{{Mooley} et~al.,}{{Mooley}
  et~al.}{2018a}]{Mooley2018}
{Mooley} K.~P.,  et~al., 2018a, \mn@doi [Nature] {10.1038/nature25452}, \href
  {http://adsabs.harvard.edu/abs/2018Natur.554..207M} {554, 207}

\bibitem[\protect\citeauthoryear{{Mooley} et~al.,}{{Mooley}
  et~al.}{2018b}]{Mooley2018b}
{Mooley} K.~P.,  et~al., 2018b, \mn@doi [Nature] {10.1038/s41586-018-0486-3},
  \href {http://adsabs.harvard.edu/abs/2018Natur.561..355M} {561, 355}

\bibitem[\protect\citeauthoryear{{Mooley} et~al.,}{{Mooley}
  et~al.}{2018c}]{Mooley2018c}
{Mooley} K.~P.,  et~al., 2018c, \mn@doi [Astrophys. J. Lett.]
  {10.3847/2041-8213/aaeda7}, \href
  {https://ui.adsabs.harvard.edu/abs/2018ApJ...868L..11M} {868, L11}

\bibitem[\protect\citeauthoryear{{Murguia-Berthier}, {Montes}, {Ramirez-Ruiz},
  {De Colle}  \& {Lee}}{{Murguia-Berthier} et~al.}{2014}]{Murguia-Berthier2014}
{Murguia-Berthier} A.,  {Montes} G.,  {Ramirez-Ruiz} E.,  {De Colle} F.,
  {Lee} W.~H.,  2014, \mn@doi [Astrophys. J.] {10.1088/2041-8205/788/1/L8},
  \href {http://adsabs.harvard.edu/abs/2014ApJ...788L...8M} {788, L8}

\bibitem[\protect\citeauthoryear{{Murguia-Berthier} et~al.,}{{Murguia-Berthier}
  et~al.}{2016}]{Murguia-Berthier2016}
{Murguia-Berthier} A.,  et~al., 2016, \mn@doi [Astrophys. J. Lett.]
  {10.3847/2041-8213/aa5b9e}, \href
  {http://adsabs.harvard.edu/abs/2017ApJ...835L..34M} {835, L34}

\bibitem[\protect\citeauthoryear{{Nagakura}, {Hotokezaka}, {Sekiguchi},
  {Shibata}  \& {Ioka}}{{Nagakura} et~al.}{2014}]{Nagakura2014}
{Nagakura} H.,  {Hotokezaka} K.,  {Sekiguchi} Y.,  {Shibata} M.,   {Ioka} K.,
  2014, \mn@doi [Astrophys. J.] {10.1088/2041-8205/784/2/L28}, \href
  {http://adsabs.harvard.edu/abs/2014ApJ...784L..28N} {784, L28}

\bibitem[\protect\citeauthoryear{{Nathanail}, {Strantzalis}  \&
  {Contopoulos}}{{Nathanail} et~al.}{2016}]{Nathanail2016}
{Nathanail} A.,  {Strantzalis} A.,   {Contopoulos} I.,  2016, \mn@doi [Mon.
  Not. R. Astron. Soc.] {10.1093/mnras/stv2558}, \href
  {http://adsabs.harvard.edu/abs/2016MNRAS.455.4479N} {455, 4479}

\bibitem[\protect\citeauthoryear{{Nathanail}, {Porth}  \&
  {Rezzolla}}{{Nathanail} et~al.}{2019}]{Nathanail2018c}
{Nathanail} A.,  {Porth} O.,   {Rezzolla} L.,  2019, \mn@doi [Astrophys. J.
  Lett] {10.3847/2041-8213/aaf73a}, \href
  {https://ui.adsabs.harvard.edu/abs/2019ApJ...870L..20N} {870, L20}

\bibitem[\protect\citeauthoryear{{Porth}, {Olivares}, {Mizuno}, {Younsi},
  {Rezzolla}, {Moscibrodzka}, {Falcke}  \& {Kramer}}{{Porth}
  et~al.}{2017}]{Porth2017}
{Porth} O.,  {Olivares} H.,  {Mizuno} Y.,  {Younsi} Z.,  {Rezzolla} L.,
  {Moscibrodzka} M.,  {Falcke} H.,   {Kramer} M.,  2017, \mn@doi [Computational
  Astrophysics and Cosmology] {10.1186/s40668-017-0020-2}, 4, 1

\bibitem[\protect\citeauthoryear{{Radice}, {Galeazzi}, {Lippuner}, {Roberts},
  {Ott}  \& {Rezzolla}}{{Radice} et~al.}{2016}]{Radice2016}
{Radice} D.,  {Galeazzi} F.,  {Lippuner} J.,  {Roberts} L.~F.,  {Ott} C.~D.,
  {Rezzolla} L.,  2016, \mn@doi [Mon. Not. R. Astron. Soc.]
  {10.1093/mnras/stw1227}, \href
  {http://adsabs.harvard.edu/abs/2016MNRAS.tmp..894R} {460, 3255}

\bibitem[\protect\citeauthoryear{{Rezzolla} \& {Zanotti}}{{Rezzolla} \&
  {Zanotti}}{2013}]{Rezzolla_book:2013}
{Rezzolla} L.,  {Zanotti} O.,  2013, Relativistic Hydrodynamics.
Oxford University Press, Oxford, UK,
  \mn@doi{10.1093/acprof:oso/9780198528906.001.0001}

\bibitem[\protect\citeauthoryear{{Rezzolla}, {Giacomazzo}, {Baiotti}, {Granot},
  {Kouveliotou}  \& {Aloy}}{{Rezzolla} et~al.}{2011}]{Rezzolla:2011}
{Rezzolla} L.,  {Giacomazzo} B.,  {Baiotti} L.,  {Granot} J.,  {Kouveliotou}
  C.,   {Aloy} M.~A.,  2011, \mn@doi [Astrophys. J. Letters]
  {10.1088/2041-8205/732/1/L6}, \href
  {http://adsabs.harvard.edu/abs/2011ApJ...732L...6R} {732, L6}

\bibitem[\protect\citeauthoryear{{Ruiz}, {Lang}, {Paschalidis}  \&
  {Shapiro}}{{Ruiz} et~al.}{2016}]{Ruiz2016}
{Ruiz} M.,  {Lang} R.~N.,  {Paschalidis} V.,   {Shapiro} S.~L.,  2016, \mn@doi
  [Astrophys. J. Lett.] {10.3847/2041-8205/824/1/L6}, \href
  {http://adsabs.harvard.edu/abs/2016ApJ...824L...6R} {824, L6}

\bibitem[\protect\citeauthoryear{{Sari}, {Piran}  \& {Narayan}}{{Sari}
  et~al.}{1998}]{Sari1998}
{Sari} R.,  {Piran} T.,   {Narayan} R.,  1998, \mn@doi [Astrophys. J. Lett.]
  {10.1086/311269}, \href
  {https://ui.adsabs.harvard.edu/abs/1998ApJ...497L..17S} {497, L17}

\bibitem[\protect\citeauthoryear{{Savchenko} et~al.,}{{Savchenko}
  et~al.}{2017}]{Savchenko2017}
{Savchenko} V.,  et~al., 2017, \mn@doi [Astrophys. J. Letters]
  {10.3847/2041-8213/aa8f94}, \href
  {http://adsabs.harvard.edu/abs/2017ApJ...848L..15S} {848, L15}

\bibitem[\protect\citeauthoryear{{Sekiguchi}, {Kiuchi}, {Kyutoku}, {Shibata}
  \& {Taniguchi}}{{Sekiguchi} et~al.}{2016}]{Sekiguchi2016}
{Sekiguchi} Y.,  {Kiuchi} K.,  {Kyutoku} K.,  {Shibata} M.,   {Taniguchi} K.,
  2016, \mn@doi [Phys. Rev. D] {10.1103/PhysRevD.93.124046}, \href
  {http://adsabs.harvard.edu/abs/2016PhRvD..93l4046S} {93, 124046}

\bibitem[\protect\citeauthoryear{Tchekhovskoy, McKinney  \&
  Narayan}{Tchekhovskoy et~al.}{2008}]{Tchekhovskoy2008}
Tchekhovskoy A.,  McKinney J.~C.,   Narayan R.,  2008, \mn@doi [Mon. Not. R.
  Astron. Soc.] {10.1111/j.1365-2966.2008.13425.x}, 388, 551

\bibitem[\protect\citeauthoryear{{The LIGO Scientific Collaboration} \& {The
  Virgo Collaboration}}{{The LIGO Scientific Collaboration} \& {The Virgo
  Collaboration}}{2017}]{Abbott2017}
{The LIGO Scientific Collaboration} {The Virgo Collaboration} 2017, \mn@doi
  [Phys. Rev. Lett.] {10.1103/PhysRevLett.119.161101}, \href
  {https://ui.adsabs.harvard.edu/abs/2017PhRvL.119p1101A} {119, 161101}

\bibitem[\protect\citeauthoryear{{The LIGO Scientific Collaboration}
  et~al.,}{{The LIGO Scientific Collaboration} et~al.}{2017}]{Abbott2017b}
{The LIGO Scientific Collaboration} et~al., 2017, \mn@doi [Astrophys. J. Lett.]
  {10.3847/2041-8213/aa91c9}, \href
  {https://ui.adsabs.harvard.edu/abs/2017ApJ...848L..12A} {848, L12}

\bibitem[\protect\citeauthoryear{{Troja} et~al.,}{{Troja}
  et~al.}{2017}]{Troja+17}
{Troja} E.,  et~al., 2017, \mn@doi [\nat] {10.1038/nature24290}, \href
  {http://adsabs.harvard.edu/abs/2017Natur.551...71T} {551, 71}

\bibitem[\protect\citeauthoryear{{Troja} et~al.,}{{Troja}
  et~al.}{2018}]{Troja2018}
{Troja} E.,  et~al., 2018, \mn@doi [Mon. Not. R. Astron. Soc.]
  {10.1093/mnrasl/sly061}, \href
  {http://adsabs.harvard.edu/abs/2018MNRAS.478L..18T} {478, L18}

\bibitem[\protect\citeauthoryear{{Troja} et~al.,}{{Troja}
  et~al.}{2019}]{Troja2019}
{Troja} E.,  et~al., 2019, \mn@doi [Mon. Not. R. Astron. Soc.]
  {10.1093/mnras/stz2248}, \href
  {https://ui.adsabs.harvard.edu/abs/2019MNRAS.489.1919T} {489, 1919}

\makeatother
\end{thebibliography}



\appendix

\section{Numerical setup and MHD models}
\label{appen}

%
\begin{table*}
  \centering
  \begin{tabular}{|l||c|c|c|c|c|c|c|c|c|c|c|c|c|c|}
    \hline
    \hline
    \!\!\!  model&$L$&$t_{\mathrm{inj}}$& $\Gamma_{\mathrm{init}}$&$\theta_{\mathrm{jet}}$&$E_{B_{\phi}}$&
    $ E_{B_{\mathrm{p}}}$& 
    $\frac{E_{B_{\mathrm{p}}}}{E_{B_{\phi}}}$&$\sigma_{\mathrm{max}}$& 
\!\!\!$\beta_{\mathrm{min}}$&$\rho_{\mathrm{max}}$& $a$ & $M_{\mathrm{tot}}$ & $M_{\mathrm{ej}}$ &$\frac{M_{\mathrm{ej}}}{M_{\mathrm{tot}}}$ \\
    \!\!\! &$[\mathrm{erg/s}]$ & $[\mathrm{s}]$ && ${\rm [deg]}$  & $[\mathrm{erg}]$ & $[\mathrm{erg}]$  &
     &&&\!\!\! $[\mathrm{g/cm^3}]$ & & $[M_{\odot}]$& $[M_{\odot}]$& $\%$\\
     &&&&&$10^{49}$&$10^{49}$&&&&$10^{10}$&&&&\\
  \hline
      \! \texttt{HD-tht.6}&$10^{51}$&$0.1$&$10$&
      $6$&$-$&$-$&$-$&$-$&$-$&$1.5 $&
      $0.9375$&$0.108$ &$0.0001$&$0.12$ \\ 
    \! \texttt{HD-tht.3}&$10^{51}$&$0.1$&$10$&
    $3$&$-$&$-$&$-$&$-$&$-$&$1.5 $&
    $0.9375$&$0.108$ &$0.0001$&$0.13$ \\
\! \texttt{MHD-p2t.03} &$-$&$-$&$-$&$-$&$5.0$   & $1.6$  & $0.3$& $0.065$ & $0.13$ & $1.5$ & $0.9375$ & $0.108$& $0.039$& $36.0$ \\
\! \texttt{MHD-p2t.03-LB}&$-$&$-$&$-$&$-$ & $0.36$ &$0.28$ & $0.3$ & $0.0026$ & $3.20$ & $2.0$ & $0.9375$ & $0.144$& $0.021$& $1.45$ \\
\! \texttt{MHD-p2t.02} &$-$&$-$&$-$&$-$&$10$ & $2.1$ & $0.2$& $0.065$ & $0.13$ & $2.0$ & $0.9375$ & $0.144$& $0.053$&$37.1$\\
\! \texttt{MHD-p2t.02-LB}&$-$&$-$&$-$&$-$ &$0.4$ & $0.084$ & $0.2$ & $0.002$ & $3.25$ & $2.0$ & $0.9375$ & $0.144$& $0.002$&$1.60$\\
\! \texttt{MHD-p2t.12} &$-$&$-$&$-$&$-$&$1.2$ & $1.5$ & $1.2$& $0.036$ & $0.13$ & $1.5$ & $0.9375$ & $0.108$& $0.036$&$34.1$  \\
\! \texttt{MHD-p2t.04}&$-$&$-$&$-$&$-$ &$4.1$ & $1.6$ & $0.4$ & $0.065$ & $0.13$ & $1.5$ & $0.9375$ & $0.108$& $0.033$&$31.2$ \\
\! \texttt{MHD-a.8-LB}&$-$&$-$&$-$&$-$  &$0.19$ & $0.115$ & $0.6$ & $0.0024$ & $3.30$ & $3.0$ & $0.8$ & $0.118$& $0.0034$&$2.10$ \\
\! \texttt{MHD-a.8-MB}&$-$&$-$&$-$&$-$ &$1.7$ & $1.05$ & $0.6$ & $ 0.02$ & $0.36$ & $3.0$ & $0.8$ & $0.118$& $0.014$&$12.0$ \\
\! \texttt{MHD-a.8}&$-$&$-$&$-$&$-$ &$3.9$ & $2.4$ & $0.6$ & $ 0.06$ & $0.13$ & $2.5$ & $0.8$ & $0.098$& $0.029$&$29.8$ \\	
\! \texttt{MHD-rout-52.4}&$-$&$-$&$-$&$-$ &$1.0$ & $0.195$ & $0.2$ & $0.0016$ & $4.10$ & $10$&$0.9375$&$0.121$&$0.018$&$15.6$ \\    
\! \texttt{MHD-600km}&$-$&$-$&$-$&$-$ &$1.7$ & $0.54$ & $0.3$ & $0.016	$ & $0.52$ & $2.0$ & $0.9375$ & $0.127$& $0.0077$&$6.23$ \\
\! \texttt{MHD-900km}&$-$&$-$&$-$&$-$ &$1.3$ & $0.4$ & $0.3$ & $0.016  $ & $0.52$ & $1.5$ & $0.9375$ & $0.106$& $0.004$&$3.94$ \\
   \hline
    \hline
  \end{tabular}  
  \caption{Properties of the various HD and MHD jets considered:
    luminosity of the HD jet ($L$), duration of the HD injection
    ($t_{\mathrm{inj}}$), initial Lorentz factor of the HD jet
    ($\Gamma_{\mathrm{init}}$), initial opening angle of the HD jet
    ($\theta_{\rm jet}$), toroidal and poloidal magnetic energies
    ($E_{B_{\phi}}, E_{B_{\mathrm{p}}}$) and their ratio, maximum
    magnetization in the torus ($\sigma:=B^2/4\pi \rho$), minimum plasma
    parameter in torus ($\beta:=p/p_m$, where $p$ and $p_m$ are the fluid
    and magnetic pressures respectively), maximum density of the torus
    ($\rho_{\mathrm{max}}$) and dimensionless spin parameter of the BH
    ($a:=J/M^2$), initial total mass ($M_{\mathrm{tot}}$), ejected mass
    ($M_{\mathrm{ej}}$) and their ratio. For all models the initial torus
    parameters are $r_{\rm in}=23.8\, \rm km$, $r_{\rm out}= 56.8\, \rm
    km$ and the matter distribution has a radial extent till $r_{\rm
      ext}=1,200\,\mathrm{km}$, whereas model \texttt{MHD-rout-52.4} has
    $r_{\rm out}= 52.4\, \rm km$, model \texttt{MHD-600km} has $r_{\rm
      ext}= 600\, \rm km$ and \texttt{MHD-900km} has $r_{\rm ext}= 900\,
    \rm km$. Note that models ending with MB and LB refer to matter with
    a medium and low magnetic-field strength, respectively, while all the
    other quantities are held the same.}
  \label{tab:appen}
\end{table*}

In this Appendix we provide details of the numerical setup of our
simulations and further show results for an extensive selection of MHD
models in order to check the robustness of the results. As anticipated,
we use \bhac to solve the general-relativistic MHD equations in a Kerr
background spacetime \citep{Porth2017}. To mimic the post-merger remnant
in GW170817 and as initial condition for the launching of an MHD jet, we
consider a non self-gravitating torus \citep{Fishbone76,Abramowicz78}
around a BH of mass $M=2.7\,M_{\odot}$ and various dimensionless spins
(see Table~\ref{tab:initial}). The radial extent of the initial matter
distribution is set to be $1,200\,\mathrm{km}$, in order to account for
the expansion of the torus, and also for the matter expelled during
merger, which has reached such a distance. To accommodate such a large
extension of matter, the numerical domain has always a radius of
$10,000\,\mathrm{km}$. Since we here focus on the production and launch
of a jet, at the beginning of the simulation all matter is bound and set
to have a zero velocity. However, we do measure the mass that becomes
unbound as a result of the jet launching and compute its contribution to
the kilonova at the end of the simulation. The simulations are performed
in two spatial dimensions using a spherical polar coordinate system. The
computational domain is resolved with either $1024\times512$ or
$512\times256$ cells and with three refinement levels, thus yielding an
effective resolution of $4092\times2048$ cells.

Over the past several years, a robust picture has been drawn on the
distribution of the ejected matter after the merger. More specifically,
BNS merger simulations indicate that the polar region is not entirely
empty of matter
\citep{Sekiguchi2016,Foucart2016a,Radice2016,Bovard2017,Dietrich2017,Fujibayashi2017b}. To
reproduce such conditions, we fill the polar region with matter, having
density that is $2.5$ orders of magnitude less than the maximum density
of the torus and a radial profile that scales like $r^{-1.5}$, with an
exception for model \texttt{HD-tht.6}, where the matter in the polar
region has $1$ order of magnitude higher density, but has the same radial
profile.  In a typical BNS merger, the two stars have a mildly strong
initial magnetic field, which is expected to be amplified during merger,
either via the Kelvin-Helmholtz or the magnetorotational instability,
yielding a very magnetic energy $>10^{50}\,{\rm erg}$, and with ratio
between poloidal and the toroidal components that is $\approx 0.3$
\citep{Kiuchi2017}. To reproduce the enhancement in the magnetic field
after the merger, we initialize our simulations with a poloidal
nested-loop magnetic field structure and a toroidal component that traces
the fluid pressure; by tuning the strength of two components it is then
possible to obtain the desired ratio in the corresponding magnetic
energies.

\begin{figure*}
  \begin{center}
      \includegraphics[width=0.43\textwidth]{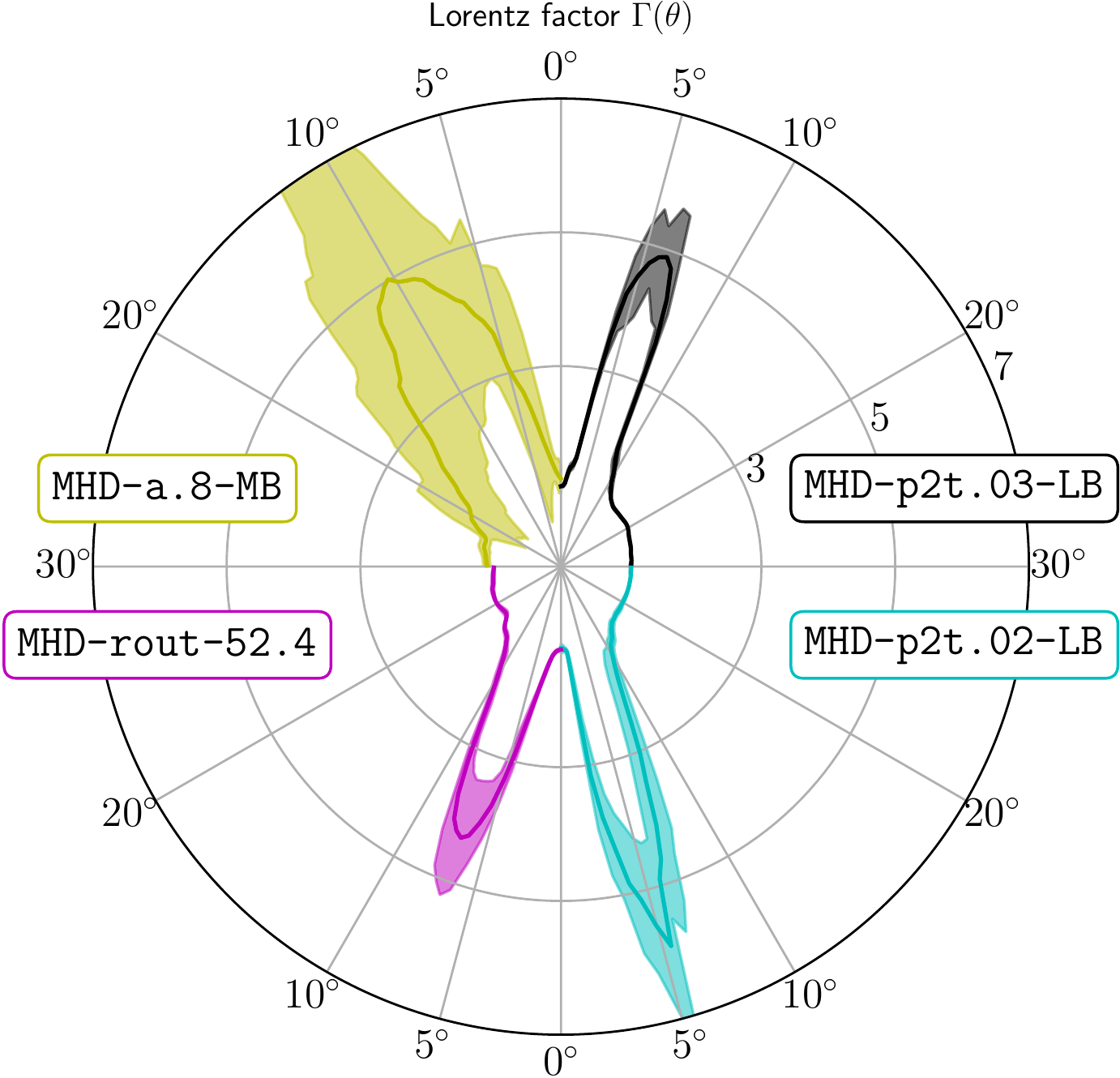}
    \hskip 1.0cm
    \includegraphics[width=0.4\textwidth]{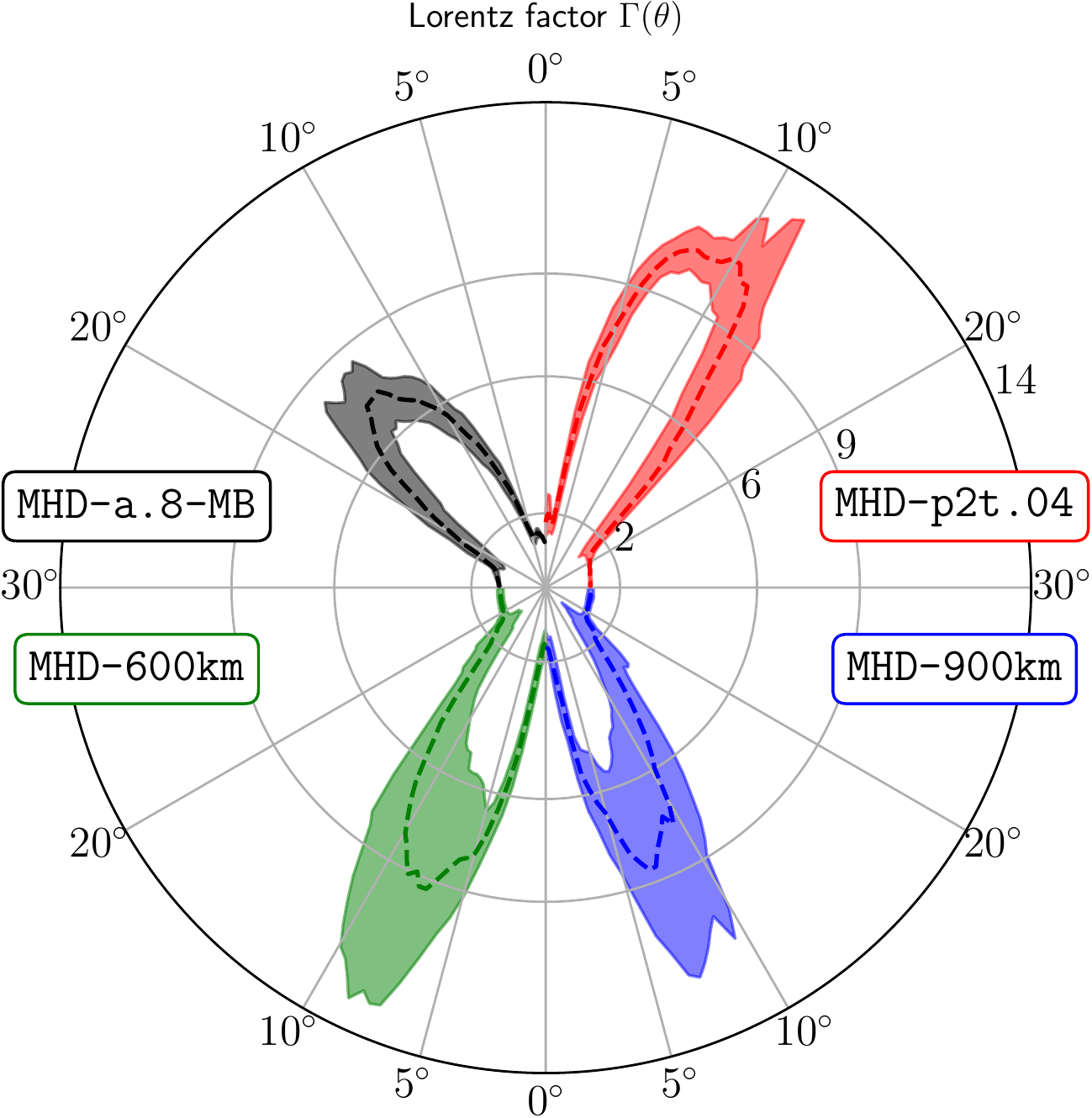}
	\end{center}
 \caption{Polar plots of the Lorentz factor for eight outflows from
   Table~\ref{tab:appen} within a cone of $30^{\circ}$, the thick lines
   show the average values, while the shaded region the $1$-$\sigma$
   variance.}
 	\label{fig:appen}
\end{figure*}

To explore a space of parameters that is as wide as reasonably possible,
we vary the initial magnetic field, the ratio of the poloidal-to-toroidal
magnetic-field energy, the spin of the BH, as well as the size and
morphology of the torus (which is ultimately dictated by the spin of the
BH). The details of all the models used are listed in
Table~\ref{tab:appen}.  For illustrative purposes, we report in
Fig.~\ref{fig:appen} the angular structure of eight outflows from
Table~\ref{tab:appen}, showing the Lorentz factor within an angle of
$0\leq\theta\leq30^{\circ}$. Similar to Fig.~\ref{fig:ang}, the Lorentz
factor (thick line) is measured in slices of constant radius , \ie
$r\sim2000\, \mathrm{km}$, and integrated over a time interval of
$\tau_{\mathrm{avg}}\sim2\, \mathrm{ms}$, the shaded areas show the
$1$-$\sigma$ variance over the time interval $\tau_{\mathrm{avg}}$, \ie
the $68\%$ variation of the Lorentz factor at each angle. From the two
polar plots it is evident that the presence of a hollow core with an
opening of $\approx 4^{\circ}-5^{\circ}$ is robust in all of the MHD
models considered in our study.

\end{document}